\def\teq#1{$\, #1\,$}                         
\def\dover#1#2{\hbox{${{\displaystyle#1 \vphantom{(} }\over{
   \displaystyle #2 \vphantom{(} }}$}}
             \font\sixrm=cmr6       
       \font\sevenrm=cmr7          

\def\erg{\varepsilon}
\def\fsc{\alpha_{\hbox{\sevenrm f}}}                                
\def\lambar{\lambda\llap {--}}
\def\rns{R_{\hbox{\sevenrm NS}}}                                
\def\emax{\erg_{\hbox{\sixrm MAX}}} 
{\catcode`\@=11 
\gdef\SchlangeUnter#1#2{\lower2pt\vbox{\baselineskip 0pt\lineskip0pt 
\ialign{$\m@th#1\hfil##\hfil$\crcr#2\crcr\sim\crcr}}}} 
\def\gtrsim{\mathrel{\mathpalette\SchlangeUnter>}} 
\def\lesssim{\mathrel{\mathpalette\SchlangeUnter<}} 
 
\def\prl{Phys. Rev. Letters} 
\def\prd{Phys. Rev. D} 
\def\rmp{Rev. Mod. Phys.}  
\def\nat{Nature}  
\def\apj{ApJ}  
\def\apjl{ApJ Lett.}  
\def\apjs{ApJ Supp.}  
\def\apss{Astr. Space Sci.}  
\def\mnras{MNRAS}  
\def\app{Astroparticle Phys.}                   
\def\asr{Adv. Space Res.}                       
\def\ass{Astrophys. Space Sci.}                       
\def\ssr{Space Sci. Rev.}                       
\documentclass[a4paper,12pt,oneside,fleqn]{article}
\usepackage{CASYS}
   \usepackage{graphicx}
\title{Photon Splitting and Pair Conversion\\ in Strong Magnetic Fields}
\author{Matthew G. Baring \\
Rice University,
Department of Physics and Astronomy - MS 108,\\
P. O. Box 1892,
Houston, Texas 77251-1892, USA\\
{\it baring@rice.edu}}

\begin{document}
\newcommand{\vol}[2]{$\,$\rm #1\rm , #2.}  
\maketitle
\begin{flushright}
\phantom{p}
\vspace{-215pt}
        To appear in Proc. CASYS '07 Conference
     ``Computing Anticipatory Systems,''\\
     eds. D. Dubois, et al. (AIP Conf. Proc., New York, 2008).
\vspace{155pt}
\end{flushright}

\begin{abstract}
The magnetospheres of neutron stars provide a valuable testing ground
for as-yet unverified theoretical predictions of quantum electrodynamics
(QED) in strong electromagnetic fields.  Exhibiting magnetic field
strengths well in excess of a TeraGauss, such compact astrophysical
environments permit the action of exotic mechanisms that are forbidden
by symmetries in field-free regions.  Foremost among these processes are
single-photon pair creation, where a photon converts to an
electron-positron pair, and magnetic photon splitting, where a single
photon divides into two of lesser energy via the coupling to the
external field.  The pair conversion process is exponentially small in
weak fields, and provides the leading order contribution to
vacuum polarization.  In contrast, photon splitting possesses no energy
threshold and can operate in kinematic regimes where the lower order
pair conversion is energetically forbidden.  This paper outlines some of
the key physical aspects of these processes, and highlights their
manifestation in neutron star magnetospheres.  Anticipated
observational signatures include profound absorption turnovers in 
pulsar spectra at gamma-ray wavelengths.  The shapes of these turnovers 
provide diagnostics on the possible action of pair creation and the
geometrical locale of the photon emission region. 
There is real potential for the first confirmation of strong field QED with the
new GLAST mission, to be launched by NASA in 2008.
Suppression of pair creation by photon splitting and its implications for
pulsars is also discussed.

\vskip 2pt
\kw: Quantum Electrodynamics, Magnetic Fields, Neutron Stars, Pulsars.
\end{abstract}

\section{Introduction}
 \label{sec:intro}
Quantum Electrodynamics (QED) is one of the most robust theories of
physics.  It has been extensively and almost exhaustively tested in 
terrestrial laboratory experiments, as have its constituent, ingredient
disciplines, special relativity and quantum mechanics.  The thoroughness
of its investigation has forged an extreme confidence in the physics
community in the viability of its predictions, and its broad applicability
in Nature.  This has naturally extended to astrophysical settings, where
both the more conventional and even the more exotic processes from
QED are applied in models of different sources of
radiation.  Foremost among these are Compton scattering
\teq{e\gamma\to e\gamma} and two-photon pair creation,
\teq{\gamma\gamma\to e^+e^-}, which have been precisely calculated
and tested since the 1930s as processes that are
probable in the cosmos due to their comparatively high cross sections.  
Because they most generally apply to
relativistic environs where X-rays and gamma-rays abound, they 
are ubiquitous in models of compact, non-magnetic astrophysical 
systems, particularly those involving black holes, both galactic (typically
with masses \teq{M} of a few solar masses \teq{M_{\odot}}), and
extragalactic and supermassive (with \teq{M\sim 10^6-10^9M_{\odot}}).

Neutron stars, compact remnants of aged stars more massive than the
sun that have undergone a supernova event, provide a very different
environment for physical processes.  Supported only by neutron
degeneracy pressure against the incredibly compressing pull of gravity, 
they stably exist with radii \teq{R_{\rm NS}} of the order of 10 km.  During the
supernova core implosion, it is believed that charge currents in their outer layers
persist and strengthen so that their magnetic fields intensify,
perhaps exceeding magnetic flux conservation.  They can then
possess fields in the range of \teq{10^{10}-10^{15}}Gauss
(i.e., \teq{10^{6}-10^{11}}Tesla; see, e.g. \cite{hnw64,st83}) 
that are far beyond the realm of terrestrial experiments.  The prime 
observational manifestations of neutron stars are pulsars, rapidly
pulsating radio sources that were discovered in 1967 \cite{hew68}.
Now seen in other wavebands such as the optical, X-rays and gamma-rays,
their interpretation as rapidly rotating neutron stars \cite{gold68} that slowly
{\it spin down} to longer pulse periods has provided powerful evidence for 
the existence of strong {\bf B} fields.  This evidence is derived from a 
rotating, tilted dipole model for the field structure \cite{sturr71,rs75},
the rotational/electromagnetic torque on which leads to inferences of 
surface fields in the range \teq{B\sim 10^{9}-10^{15}}Gauss.  Such a dipole
picture is the simplest choice for the field geometry, being the leading order
moment for any magnetic field configuration.  The rotation and magnetic axes are 
generally not aligned, just as in the cases of the sun and the Earth, and so
the time-varying {\bf B} fields induce electric fields {\bf E} of substantial 
magnitude above the stellar surface.  These {\bf E} fields are
quenched \cite{sturr71,rs75} on relatively small lengthscales by charges that are
prolifically created via the magnetic pair production process that is 
detailed in this paper.   Furthermore, note that additional support
for the existence of such intense {\bf B} fields comes from spectroscopic
observations of putative cyclotron emission or absorption lines in the 
20--150 keV band of accreting X-ray binary pulsars \cite{trumper78}.
A discussion of these pertinent astrophysical elements can be 
found in \cite{st83}.

These strong field environs provide an opportunity for more exotic and
less familiar predictions of QED to come to the fore.  The {\bf B} field
profoundly changes the physical nature of quantum interactions, yielding
a strong bias towards momentum exchange along the field.
It quantizes the electron states in a direction perpendicular to the field so
that momentum in that direction is no longer conserved; a product of
the lack of translational invariance of the states orthogonal to {\bf B}.
Accordingly, \teq{B=0} symmetries are broken, and interactions that are 
forbidden in field-free environs
become possible, and even probable in neutron star magnetospheres.
The most likely among these are magnetic pair creation \teq{\gamma\to e^+e^-}
and photon splitting \teq{\gamma\to\gamma\gamma}.  These are more
recent predictions of QED, dating from the 1950s and 1960s, partly due
to their inherent mathematical complexity (spawned by quantization in a cylindrical
geometry), and partly due to the contemporaneous
rising interest in their applicability to cosmic systems, e.g. neutron stars.
The fact that these, as yet, untested predictions of QED are widely invoked
in pulsar contexts is a true testament to physicists' confidence in the theory
of relativistic quantum mechanics.  These two processes form the focus
of this paper, though it is noted that strong-field quantum effects also have profound
influences on Compton scattering \cite{dh86,bam86}, classical synchrotron 
radiation \cite{st68,baring88}, and other radiation processes.
  
\section{Magnetic Pair Creation}
 \label{sec:pprod}
One-photon pair production \teq{\gamma\to e^+e^-} in strong
magnetic fields is a first-order QED process with one vertex that is quite
familiar to pulsar theorists, having been invoked in polar cap models to
explain the photon emission 
of both radio pulsars (e.g. \cite{sturr71,rs75})
and the handful of known gamma-ray pulsars (e.g. \cite{dh82}).  
It is forbidden in field-free regions due
to the imposition of four-momentum conservation, but takes place in an
external magnetic field, which can absorb momentum perpendicular to 
{\bf B} (momentum along the field is still conserved, as is the energy).  The first analytic computations of its rate \teq{R^{\rm pp}} \cite{toll52,klep54} (hereafter the superscript {\bf pp}
denotes pair production) indicated a rapid rise with
increasing photon energy \teq{\erg_{\gamma}} and magnetic field strength, becoming
significant for \teq{\gamma}-rays above the pair threshold, \teq{\omega \equiv 
\erg_{\gamma}/(m_ec^2) = 2/\sin\theta_{\rm kB}}, and for fields approaching 
the quantum critical field 
\begin{equation}
   B_{\rm cr} \; =\; \dover{m_e^2c^3}{e\hbar}
	\;\approx\; 4.413 \times 10^{13} \;\hbox{Gauss}\quad .
 \label{eq:Bcrit}
\end{equation}
Here \teq{\theta_{\rm kB}} is the angle the
photon momentum vector {\bf k} makes with the magnetic field {\bf B}.  \teq{B_{\rm cr}}
represents the field at which the cyclotron energy equals \teq{m_ec^2}, and
defines the field scale at which the impact of the external {\bf B} field on quantum 
processes becomes significant.   It is appropriate to identify now the dimensionless
scaling conventions to be adopted throughout this paper that are of 
common usage.  All photon energies, \teq{\omega},
will be scaled in terms of the electron rest mass energy \teq{m_ec^2}, and all
field strengths will hereafter be expressed in terms of \teq{B_{\rm cr}}, i.e.
\teq{B=1} denotes a field of \teq{4.413\times 10^{13}}Gauss.  We note in passing that pair creation 
\teq{\gamma\to e^+e^-} in the induced electric fields is also possible in
rotating pulsar magnetospheres, however this is reduced by the order of
\teq{v_{\rm rot}/c\ll 1} near the neutron star surface due to corotation speeds
\teq{v_{\rm rot}=2\pi \rns /P} usually being relatively small (\teq{\rns} is the neutron 
star radius and \teq{P} is the pulsar period).

The magnetic pair creation rate is resonant at the thresholds for each combination
of produced pair states, due to the available momentum parallel to the field approaching
zero in the frame where \teq{\theta_{\rm kB}=\pi/2}.  
A large number of integrable (over photon energies) resonances
result, producing a characteristic sawtooth structure \cite{dh83,bk07} 
that is displayed in Figure~\ref{fig:pprod_split} below.  In astrophysical contexts,
a range of field strengths and initial photon energies \teq{\omega} are always sampled,
and the resulting convolution smears out the sawtooth appearance into a continuum.
Fully general expositions of the \teq{\gamma\to e^+e^-} rate in uniform
{\bf B} fields can be found in \cite{st68,dh83}.  Here we highlight the spectral
structure near absolute pair threshold \teq{\omega = 2/\sin\theta_{\rm kB}}, 
a case most germane to pulsar applications
since generally photons are produced in beams almost along the local field
lines in the dipole geometry, so that pair creation threshold is crossed from below
during magnetospheric propagation of photons.  This implies that
calculations in the domain \teq{\omega\gtrsim 2/\sin\theta_{\rm kB}} are typically
of greater practical relevance, especially for near-critical or supercritical
fields \teq{B\gtrsim 0.1} \cite{bh01}.  The pertinent rates, exhibited in \cite{bh01},
can be written as follows.  Let \teq{(j,k)} denotes the Landau
level quantum numbers of the produced pairs, which have energies
\teq{\sqrt{1+p_{jk}^2+2 jB}} and \teq{\sqrt{1+p_{jk}^2+2 kB}}.  The parallel
momenta \teq{p_{jk}} of the pairs in the frame of reference for photons moving 
perpendicular to the field are given by
\begin{equation}
   \Bigl\vert p_{jk}\Bigr\vert \; =\; 
   \left[ \dover{\omega^2}{4} \sin^2\theta_{\rm kB} - 1 - (j+k)B + 
   \left( \dover{(j-k)B}{\omega\sin\theta_{\rm kB}} \right)^2\right]^{1/2}  \quad ,
 \label{eq:p_par}
\end{equation}
with the solution of \teq{p_{jk}=0} defining the host of resonant energies
\teq{\omega}.  Throughout, electron energies are scaled by \teq{m_ec^2} and
their momenta are in terms of \teq{m_ec}.

Near threshold, only a few pair states are kinematically accessible.
The rates are dependent on the polarization of the
incoming photon.  Following common practice, here we adopt the convention
that photon linear polarizations are such that
\teq{\parallel} refers to the state with the photon's {\it electric}
field vector parallel to the plane containing the magnetic field and
the photon's momentum vector, while \teq{\perp} denotes when the photon's
electric field vector is normal to this plane.   Let \teq{\lambar =\hbar/m_ec}
be the Compton wavelength of the electron over \teq{2\pi}, and \teq{\fsc =e^2/\hbar c}
be the fine structure constant.  The exact, polarization-dependent, pair production 
rate \cite{dh83}, including only the \teq{(j=0,k=0)} pair state for \teq{\parallel} polarization is:
\begin{equation}
   R^{\rm pp}_{\parallel} \; =\;  \dover{\fsc c}{\lambar}\; \dover{\sin\theta_{\rm kB}}{\xi 
   |p_{\hbox{\sevenrm 00}}|}\,\exp(-\xi)\quad , \quad 
   \omega \ge \dover{2}{\sin\theta_{\rm kB}}\quad ,
  \label{eq:tpp_par}
\end{equation}
for \teq{\xi =\omega^2\sin^2\theta_{\rm kB}/[2B]}, while the sum of the \teq{(j=0,k=1)} 
and \teq{(j=1,k=0)} states contributes just above threshold
for the \teq{\perp} polarization:
\begin{equation}
   R^{\rm pp}_{\perp} \; =\; \dover{\fsc c}{\lambar}\; \dover{\sin\theta_{\rm kB}}{\xi 
   |p_{\hbox{\sevenrm 01}}|}\, \Bigl(E_0 E_1 + 1 + p_{01}^2\Bigr)\,\exp(-\xi) \quad , \quad
   \omega \ge \dover{1+\sqrt{1 + 2B}}{\sin\theta_{\rm kB}}\quad .
  \label{eq:tpp_perp}
\end{equation}
The pertinent dimensionless energies of the produced pairs are 
\begin{equation}
   E_0  \; =\; \sqrt{1 + p_{01}^2}\quad , \quad  
   E_1  \; =\; \sqrt{1 + p_{01}^2 + 2B}\quad .
 \label{eq:pair_energies}  
\end{equation}
Observe that the absolute threshold for \teq{\perp} photons is higher
than that for \teq{\parallel} photons, an effect that finds its origin in the
spin-dependence of energies in the final pair states.  Indeed, the rate
for the \teq{\perp} state is always lower than that for the \teq{\parallel} state.
This property for the leading order photon absorption process, when connected 
via the optical theorem to the
consideration of dispersion in the magnetized vacuum,
clearly indicates that such a vacuum is birefringent.

For photon energies well above
threshold, the number of accessible pair states becomes very large, 
much larger than just the few states considered above (see Figure~\ref{fig:pprod_split}
for an illustration of the sawtooth resonant structure of the rates).  Then the
resonances at \teq{p_{jk}=0} tend to blend into one another, an effect that is enhanced in pulsar 
magnetospheres by sampling 
significant ranges of photon energies, angles and field strengths that smear out the cusp-like
structure in forming an average reaction rate.  In this regime, the dependence on \teq{\omega}
can be averaged over ranges larger than the typical separation of the resonances,
and one can use more convenient asymptotic expressions for the polarization-dependent
pair creation rates \cite{klep54,erber66,te74}:
\begin{equation}
   R^{\rm pp}_{\parallel} \;\approx\; 2 R^{\rm pp}_{\perp} \;\approx\; \dover{1}{2}
    \sqrt{\dover{3}{2}}\, \dover{\fsc c}{\lambar} B \sin\theta_{\rm kB}
  \,\exp \biggl\{ -\dover{8}{3\chi} \biggr\}  \, ,\quad 
  \chi \,\equiv\, \omega B\sin\theta_{\rm kB} \,\ll\, 1\; ,
   \label{eq:pp_asymp_rate}
\end{equation}
where \teq{\chi} is the critical asymptotic expansion parameter.  These results essentially
derive from Schwinger-type mathematical formulations, as do alternative
results (e.g. \cite{erber66,te74})
for the \teq{\chi\gg 1} regime that are generally of less practical interest.
The differences between the asymptotic result in Eq.~(\ref{eq:pp_asymp_rate})
and the exact rates become profound near threshold \cite{dh83}, and the
asymptotic analyses can be improved considerably by treating mildly-relativistic
regimes for the produced pairs \cite{baring88,bk07}.

\begin{figure}[ht]
   \begin{center}
   \includegraphics[width=4.75truein]{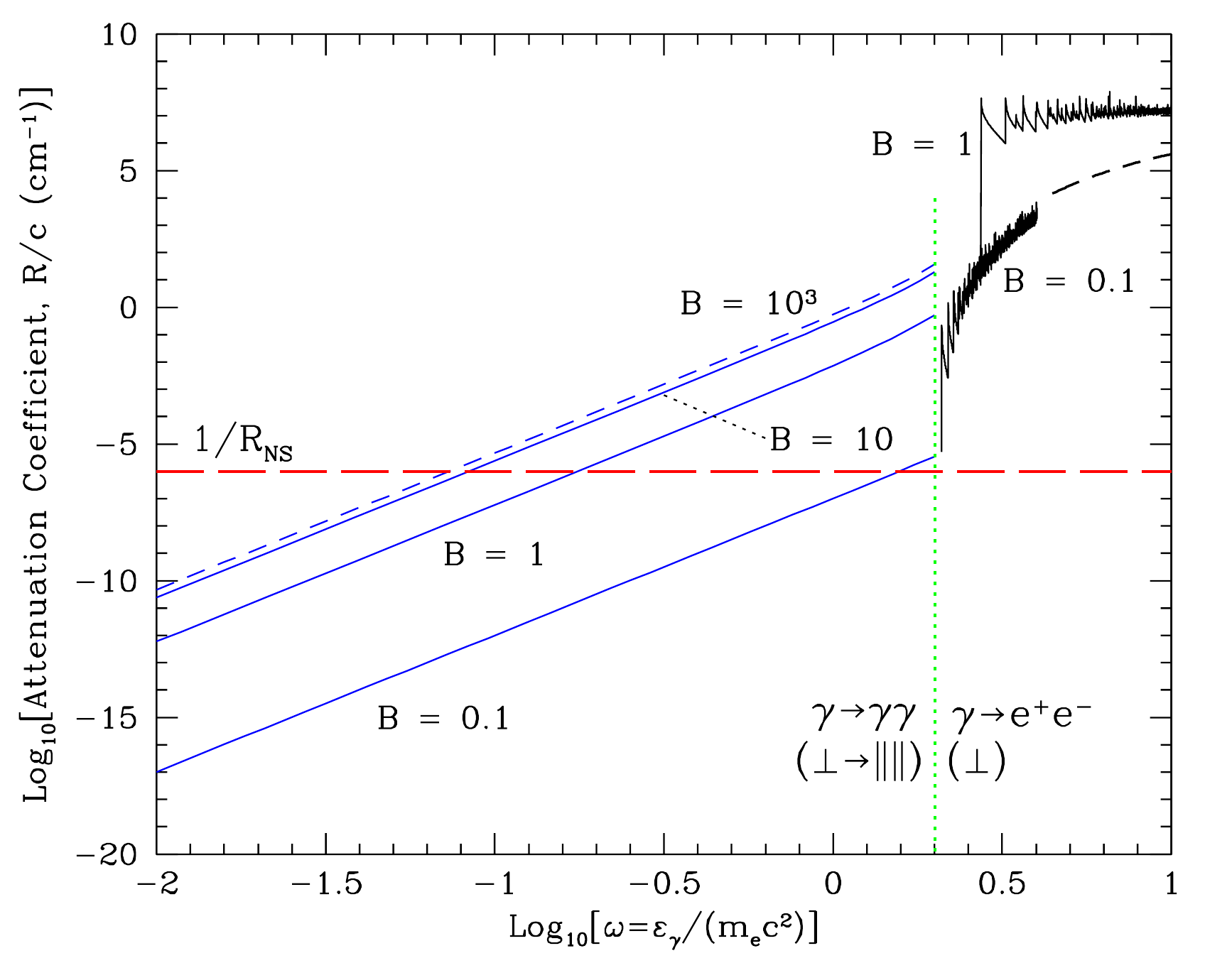}
   \end{center}
   \vspace{-20pt}
\caption{
Attenuation coefficients or inverse attenuation lengths (i.e., rates \teq{R^{\rm pp}_{\perp}} and 
\teq{R^{\rm sp}_{\perp\to\parallel\parallel}} divided by \teq{c}), for pair
creation (above the threshold \teq{\omega=2}) and photon splitting
for the polarization mode \teq{\perp\to\parallel\parallel} (only for \teq{\omega < 2}), 
as functions of the incident photon energy \teq{\omega}, for field strengths \teq{B}, 
as labelled.  Photons are assumed to propagate orthogonally to the field lines
(\teq{\theta_{\rm kB} =90^\circ}).  The pair creation rates are computed
using \cite{dh83,bh97}, and only for \teq{\perp} photons; at \teq{\omega > 4} 
the dashed \teq{B=0.1} curve
is the asymptotic form in Eq.~(\ref{eq:pp_asymp_rate}).  Also, the dashed splitting 
curve labelled \teq{B=10^3} represents the asymptotic high-field ``saturation'' result.  
The horizontal heavy dashed line labelled \teq{1/\rns} marks the approximate minimum 
effective attenuation coefficient for which these processes become prolific in pulsar
magnetospheres.
}
 \label{fig:pprod_split}
\end{figure}

As mentioned above, in polar cap pulsar models \cite{sturr71,rs75,dh82}, 
high energy radiation is usually emitted at very small angles
to the magnetic field, well below pair threshold \cite{bh01}.  The \teq{\gamma}-ray photons will convert into pairs only after
traveling a distance \teq{s} that is a fraction of the field line radius of curvature
\teq{\rho_c} (which exceeds the neutron star radius \teq{\rns}), 
so that \teq{\sin\theta_{\rm kB} \sim s/\rho_c}.  From Eq.~(\ref{eq:pp_asymp_rate}),
the pair production rate will be vanishingly small until the argument of the
exponential approaches unity, i.e., when \teq{\omega B\sin\theta_{\rm kB} \gtrsim
0.2}.  Consequently, pair production will occur well above threshold when \teq{B
\ll 0.1} and the asymptotic expression will be valid; from Figure~\ref{fig:pprod_split}
one can deduce that the attenuation length will then be much less than \teq{\rns}. 
In contrast, when \teq{B \gtrsim  0.1}, pair creation will occur at or near threshold,
and Eqs.~(\ref{eq:tpp_par}) and~(\ref{eq:tpp_perp}) become more appropriate.  
In this domain, an additional sublety
arises, since another mode of pair creation exists, namely the formation of pairs in
a bound state, i.e. positronium.  This has been proposed as an
effective competitor to the production of free pairs \cite{su85,hrw85,um95} 
because the binding energy lowers the threshold slightly
(\teq{\ll 1}\%) below the value for production of free pairs.  
The relevance of positronium formation to pulsars is potentially great
if the bound state is stable for considerable times.  Positronium is
subject to destruction by three main mechanisms: free decay, electric
field ionization,  and photo-ionization.  As discussed in \cite{bh01}, due to
conflicting calculations, it is presently unclear whether or not positronium
is stable to free decay and photo-ionization; assessment of this issue motivates
future work.

\section{Magnetic Photon Splitting}
 \label{sec:split}
The most important competitor to pair creation \teq{\gamma\to e^+e^-}
at high magnetic field strengths as a mechanism for attenuating photons
in pulsar magnetospheres is magnetic photon splitting
\teq{\gamma\to\gamma\gamma}.   The relevance of this process
to neutron star environments has been emphasized in various papers
\cite{adler71,mitro86,baring88}.  Splitting is a third-order
QED process with a triangular Feynman diagram, and therefore is weaker
than magnetic pair creation by the order of \teq{\fsc^2}, yet without the
constraint of a threshold energy, an important distinction.  Though splitting is permitted
by energy and momentum conservation, when \teq{B=0} it is forbidden by
a charge conjugation symmetry of QED known as Furry's theorem (e.g.
\cite{jr80}), which states that ring diagrams that have an
odd number of vertices with only external photon lines generate
interaction matrix elements that are identically zero.  
The presence of an external field 
breaks this symmetry.  The splitting of photons is therefore a
purely quantum effect, and has appreciable reaction rates only when
\teq{B\gtrsim 1}.

Magnetic splitting \teq{\gamma\to\gamma\gamma} is a relatively recent prediction of
QED. A decade of controversy followed the earliest calculations in the 1960s before
the first correct evaluations of its rate \cite{adler71,bb70,adler_etal70} were 
performed, via an effective Langrangian technique. These works focused on asymptotic forms in
the limit of photon energies well below pair creation threshold, which
varied as \teq{B^6} when \teq{B\ll 1}.   Their rate
determinations neglected dispersion in the birefringent, magnetized
vacuum, so that all photon momenta were assumed collinear.  The early
controversy was fueled by the inherent difficulties in calculating  the
rates of this third order process by standard QED techniques, an issue
that re-emerged in the 1990s (see \cite{bh97} for a discussion).   For weak-dispersion
regimes, fully
general rates in Schwinger-type formalisms were derived by various groups
\cite{adler71,stone79,bms96,as96}.  Even more general S-matrix, Landau representation calculations
that include information on the pair resonances are derived by
\cite{mbw94} and \cite{wbm98,baring00} 
(note that issues concerning the erroneous numerics
in \cite{mbw94} are addressed in \cite{bh97}).  These resonances
are a significant factor in determining the splitting rate near and above pair
threshold, since then the intermediate pair states go ``on-shell'' and the
splitting process effectively becomes first-order in \teq{\fsc}, tantamount
to a pair conversion followed by cyclotron decays.

It is practical to
restrict considerations of splitting to regimes of weak dispersion,
where manageable expressions for its rates are obtainable.  These are
still complicated triple integrations in Schwinger formalisms 
\cite{adler71,stone79,bms96,as96} 
or triple summations in S-matrix calculations \cite{baring00}.  Further specialization to either
low magnetic fields (\teq{B\ll B_{\rm cr}}) or low photon energies
(\teq{\omega\ll 2}) therefore proves expedient, and palatable results
for splitting rates were first obtained in such regimes.  A compact presentation 
of these rates (i.e. for \teq{\omega\ll 2}) for the three polarization modes of splitting
permitted by CP (charge-parity) invariance in QED, namely
\teq{\perp\to\parallel\parallel}, \teq{\perp\to\perp\perp} and
\teq{\parallel\to\perp\parallel}, is 
\begin{eqnarray}
   R^{\rm sp}_{\perp\to\parallel\parallel}(\omega ) &=& 
   \dover{\fsc^3}{60\pi^2}\, \dover{c}{\lambar}\, 
   \omega^5\; B^6\; \sin^6\theta_{\rm kB}\; {\cal M}_1^2 \; =\; 
   \dover{1}{2}\, R^{\rm sp}_{\parallel\to\perp\parallel}
  \nonumber\\[-5.5pt]
  &&  \label{eq:splittatten} \\[-5.5pt]
   R^{\rm sp}_{\perp\to\perp\perp} &=& 
   \dover{\fsc^3}{60\pi^2}\, \dover{c}{\lambar}\, 
   \omega^5\; B^6\; \sin^6\theta_{\rm kB}\; {\cal M}_2^2\nonumber
\end{eqnarray}
using the superscript notation {\bf sp} to signify photon splitting.  Here
the ``scattering amplitudes,'' which have been computed numerically in
\cite{bh97,adler71}, are given by
\begin{eqnarray}
   {\cal M}_1 &=& \dover{1}{B^4}
   \int^{\infty}_{0} \dover{ds}{s}\, e^{-s/B}\,
   \Biggl\{ \biggl(-\dover{3}{4s}+\dover{s}{6}\biggr)\,\dover{\cosh s}{\sinh s} 
   +\dover{3+2s^2}{12\sinh^2s}+\dover{s\cosh s}{2\sinh^3s}\Biggr\}\;\; ,
  \nonumber\\[-5.5pt]
  &&   \label{eq:splitcoeff} \\[-5.5pt]
   {\cal M}_2 &=& \dover{1}{B^4}
   \int^{\infty}_{0} \dover{ds}{s}\, e^{-s/B}\,
   \Biggl\{ \dover{3}{4s}\,\dover{\cosh s}{\sinh s} +
   \dover{3-4s^2}{4\sinh^2s} - \dover{3s^2}{2\sinh^4s}\Biggr\}\;\; .\nonumber
\end{eqnarray}
Evaluations of these forms in terms of special functions were obtained in \cite{baring00}:
\begin{eqnarray}
 {\cal M}_1 & = & \dover{1}{B^3}\;
   \Biggl\{\, \dover{4}{B} \log_e\Gamma_1\Bigl(\dover{1}{2B}\Bigr)
      - \dover{1}{2B^2} \log_e\Gamma\Bigl(\dover{1}{2B}\Bigr)
      - \biggl( \dover{1}{3B} + \dover{1}{4B^3}\biggr)
                     \psi \Bigl(\dover{1}{2B}\Bigr)  \nonumber\\[-3.5pt]
   && \quad  - \dover{1}{12B^2} \,\psi' \Bigl(\dover{1}{2B}\Bigr) 
      - \dover{1}{6} - \dover{24 L_1 -1}{6B}
      - \dover{1}{4B^2} \biggl( \log_e \dover{4 B^3}{\pi} +1 \biggr) 
        \Biggr\}   \nonumber\\[-5.5pt]
 \label{eq:M12_specfunc}\\[-5.5pt]
 {\cal M}_2 & = & \dover{1}{B^3}\;
   \Biggl\{\, - \dover{3}{2B^2} \log_e\Gamma\Bigl(\dover{1}{2B}\Bigr)
      + \dover{3}{4B^3}\, \psi \Bigl(\dover{1}{2B}\Bigr)  \nonumber\\[-3.5pt]
   && \quad   + \dover{1}{8B^4} \,\psi' \Bigl(\dover{1}{2B}\Bigr) 
      + \dover{1}{3B} + \dover{1}{2B^2} - \dover{1}{B^3}
      + \dover{3}{4B^2} \, \log_e 4\pi B  \Biggr\} \;\; ,  \nonumber
\end{eqnarray}
where \teq{\Gamma (x)} is the Gamma function, \teq{\psi (x)=d\log_e\Gamma (x)/dx}
is the digamma function, \teq{L_1\approx 0.24875} is a constant resulting from the
sum of the logarithmic series that expresses the remainder for Stirling's approximation 
for the \teq{\Gamma} function, and
\begin{equation}
  \log_e\Gamma_1(x)\; =\; \int_0^x dt\,\log_e\Gamma (t)
  + \dover{1}{2}\, x(x-1) - \dover{x}{2}\log_e2\pi\quad .
 \label{eq:Gamma1def}
\end{equation}
At low fields, \teq{{\cal M}_1\approx 26/315} and \teq{{\cal M}_2\approx 48/315} are
independent of \teq{B}, but at high fields possess \teq{{\cal M}_1\approx
1/(6B^3)} and \teq{{\cal M}_2\approx 1/(3B^4)} dependences.  Hence,
the rate for \teq{\perp\to\parallel\parallel} asymptotically approaches a constant
as \teq{B\to\infty}, a result that is exhibited in Figure~\ref{fig:pprod_split}. 
These rates are of broad applicability to
pulsar modeling.  Deviations from this low energy limit near pair
creation threshold are detailed by \cite{bh97,bms96,baring00}; they
are somewhat apparent in Figure~\ref{fig:pprod_split}, for which the numerics
were obtained from the full Schwinger-type results of \cite{bh97}.
Note that the \teq{B^6} dependence of the rates when \teq{B\ll 1} permits
the use of so-called hexagon Feynman diagrams when treating the field as
a small perturbation at three vertices, a technique employed in early computations
of this process.   This approximation cannot be used for
higher \teq{B} where the field's influence on the electron propagators must
be incorporated self-consistently, and results in greater complexity as appears
in Eq.~(\ref{eq:splitcoeff}).

As mentioned in Section~\ref{sec:pprod}, the different pair creation attenuation rates
for the two linear polarization states automatically generate polarization-dependent
refractive indices (\teq{n_{\perp} \neq n_{\parallel}}) and light speeds, i.e. birefringence.
This birefringence of the magnetized vacuum also implies an alteration of the
kinematics of strong field QED processes \cite{adler71}, admitting the
possibility of non-collinear photon splitting.  Hence, while the
splitting modes \teq{\perp\to\perp\parallel},
\teq{\parallel\to\perp\perp} and \teq{\parallel\to\parallel\parallel}
are forbidden by CP invariance in the limit of zero dispersion,
dispersive effects guarantee a small but non-zero probability for the
\teq{\perp\to\perp\parallel} channel.  Extensive discussions of linear
dispersion in a magnetized vacuum \cite{adler71,shabad75} 
demonstrate that in the limit of {\it weak linear vacuum} dispersion
(roughly delineated by $B\sin\theta_{\rm kB} \lesssim B_{\rm cr}$),
where the refractive indices for the polarization states are very close
to unity, energy and momentum could be simultaneously conserved only
for the splitting mode \teq{\perp\to\parallel\parallel} (of the modes
permitted by CP invariance) below pair production threshold.  

This result,
known as Adler's \cite{adler71} kinematic selection rules for photon splitting,
is contingent upon the inequality \teq{R^{pp}_{\perp} < R^{pp}_{\parallel}},
assuming that pair creation \teq{\gamma\to e^+e^-} provides the
dominant contribution to the vacuum dispersion, generally true for \teq{B\ll 1}
regimes of low plasma density.\footnote{ 
Note that in the dense gases encountered
deep in neutron star atmospheres, dispersion becomes dominated by plasma 
contributions \cite{bulik98}, for which the selection rules of Adler are not applicable.  
}
Therefore, it is probable that only the one mode (\teq{\perp\to\parallel\parallel}) 
of splitting operates in normal pulsars.  However, this constraint may not hold
in supercritical fields where stronger vacuum dispersion arises and
other contributions may come into play.  For example, the generalized vacuum
polarizability tensor may yield significant corrections from quadratic 
(i.e. those that connect to photon splitting) and higher order
contributions.  Recent analysis \cite{wm06} of vacuum dispersion induced only by
pair attenuation has indicated that while the selection rules are generally upheld 
as \teq{\omega} increases towards pair threshold, they have to be modified in
certain phase spaces above pair threshold.  In such regimes, splitting cannot
be discounted relative to \teq{\gamma\to e^+e^-} as a higher order process, since
its rates sample pair resonances when the electron propagators go ``on-shell,''
and the splitting rate may approach that of a first-order process.  Such calculations
of rates for \teq{\gamma\to\gamma\gamma} at \teq{\omega > 2/\sin\theta_{\rm kB}}
have not yet been performed at any length, though the formulations of
\cite{mbw94,baring00} can be readily applied to this problem.

\section{Anticipated Astrophysical Signatures}
 \label{sec:astrosignal}
Having encapsulated the essentials of the physics of these two exotic
predictions of high-field quantum electrodynamics, the focus now turns
to how signatures of these processes might realistically manifest themselves 
in pulsars.  It is quite possible that 
a vindication in astrophysical contexts for the theoretical expectations 
may predate the eventual first  observation of either process in terrestrial
laboratories.  

\subsection{Pair Creation}
 \label{sec:pprod_obs}

\begin{figure}[ht]
   \begin{center}
   \includegraphics[width=5.0truein]{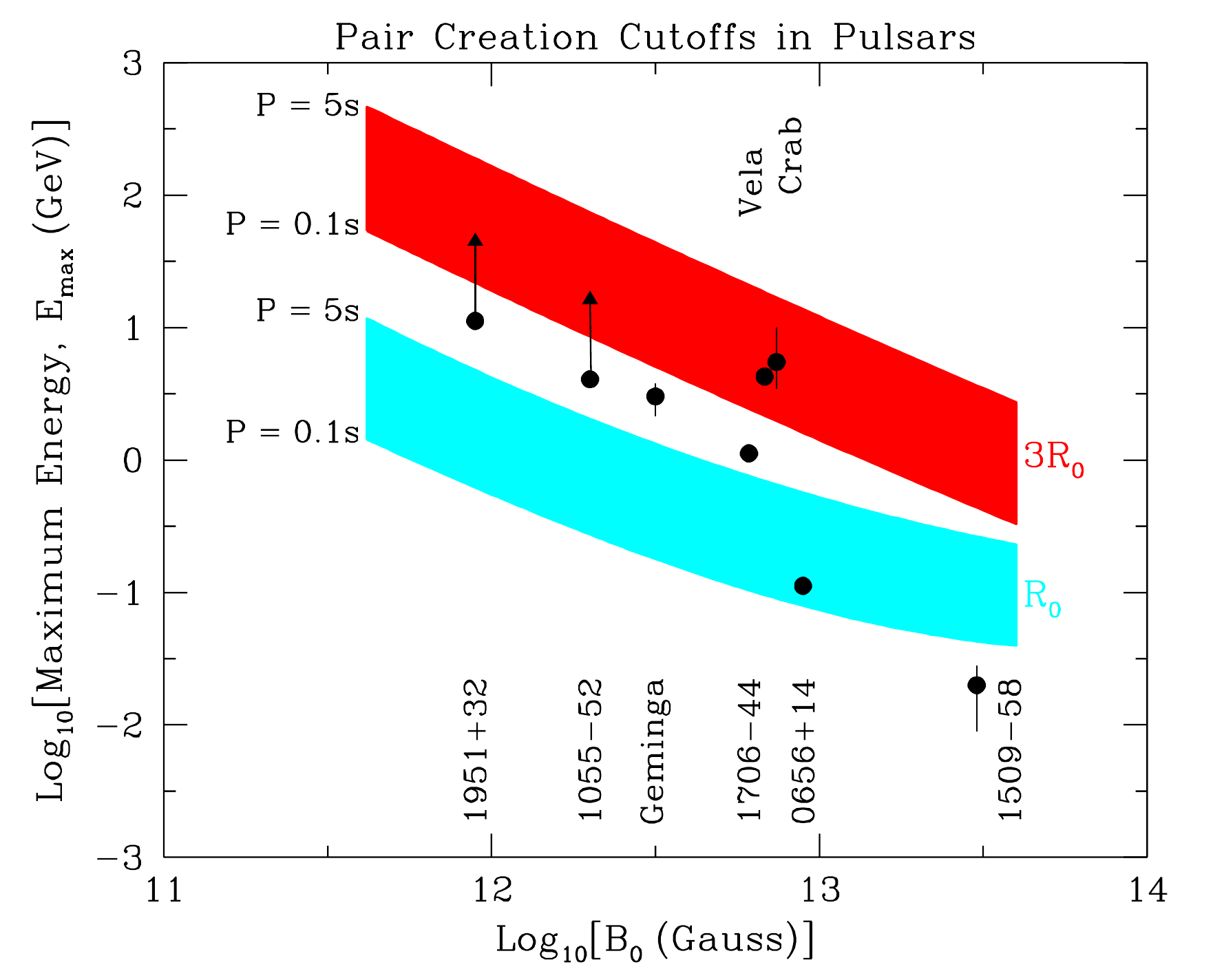}
   \end{center}
   \vspace{-25pt}
\caption{Maximum pulsar emission energies (adapted from \cite{baring04})
imposed by pair creation attenuation at two different altitudes,
\teq{R_0} (dashed curves) and \teq{3R_0} (solid curves), described
empirically via Eq.~(\ref{eq:emax}).  For each altitude, a range
of pulse periods (polar cap sizes) \teq{0.1-5}sec is represented, as indicated.  These
energies are determined by the comprehensive photon
propagation/attenuation code described in \cite{bh01},
which includes curved spacetime effects.  Observed/inferred cutoff energies (or
ranges) for 8 gamma-ray pulsars of different \teq{B_0} are indicated,
from which a trend of declining altitude of emission with increasing surface polar
field \teq{B_0} is suggested.  
}
 \label{fig:hecutoff} 
\end{figure}

Since the magnetic pair creation rate is generally a strongly increasing function
of photon energy, its major signature is an attenuation turnover in
pulsar radiation spectra.  The resulting maximum energy of emission
is controlled by attenuation during photon propagation through the pulsar
magnetosphere.  Such attenuation provides a characteristic
super-exponential turnover \cite{dh96} (see below) that
contrasts that expected in outer gap models (e.g. see \cite{thomp01,rh07} for
a comparison), scenarios where the gamma-rays are generated in a pulsar's
outer magnetosphere and the quantum effects discussed here are negligible.  
Pair creation effectively occurs at the threshold
\teq{\omega\sin\theta_{\rm kB} =2} for high fields, i.e.  \teq{B\gtrsim 1},
and above threshold at \teq{\omega\sin\theta_{\rm kB}
\sim 0.2/B} for lower fields \cite{dh83}.  Hence, the mean free path for
photon attenuation in {\it curved} fields is \teq{\lambda_{\rm pp} \sim
\rho_c/\omega\, \max \{ 2, \; 0.2/B \} }, i.e. usually when
\teq{\omega\sin\theta_{\rm kB}} crosses above threshold during
propagation.  This assertion derives from the fact that photons are generally
emitted almost parallel to the local {\bf B} due to the relativistic beaming
associated with electrons/pairs streaming along the field lines.
The radius of field curvature at altitude \teq{R_0\geq R_{\rm NS}}
is \teq{\rho_c = [P\, R_0\, c/2\pi]^{1/2}} for a pulsar period \teq{P}.  This introduces
a dependence of the energy when pair threshold is crossed of
\teq{\omega\propto1/\theta_{\rm kB}\propto \rho_c\propto \sqrt{P\, R_0}}.  In a dipole
field geometry, the local magnetic field scales as \teq{B\propto R_0^{-3}},
thereby generating the scaling \teq{\omega\propto 1/B\propto R_0^3} for the energy
of photons that access pair threshold.  Accordingly, 
the approximate dependence of pair creation cutoff energies \teq{\emax} on 
the surface polar field strength \teq{B_0}, altitude
\teq{R_0} and pulsar period \teq{P} (in seconds) can be summarized in
the approximate empirical 
relation (e.g. \cite{baring04} and references therein)
\begin{equation} 
   \emax \approx 0.4 \sqrt{P} \, \biggl( \dover{R_0}{R_{\rm NS}} \biggr)^{1/2} \; 
   \max \Biggl\{ 1,\; \dover{0.1}{B_0}\,  
   \biggl( \dover{R_0}{R_{\rm NS}} \biggr)^3 \Biggr\}\; \hbox{GeV} \;\; . 
 \label{eq:emax} 
\end{equation} 
Accurate numerical determinations of the turnover maximum energy 
from the codes developed in \cite{hbg97,bh01}, are plotted in
Figure~\ref{fig:hecutoff}; these include the effects of general
relativity on spacetime curvature, field enhancement and photon energy
in ``slowly-rotating'' systems with \teq{P\gg 2\pi R_{\rm NS}/c}.  
At fields \teq{B_0\gtrsim 0.7}
photon splitting acts to further reduce \teq{\emax}, as discussed in \cite{bh01},
though not shown in the Figure here. 

Figure~\ref{fig:hecutoff} and Eq.~(\ref{eq:emax}) clearly indicate
a strong anti-correlation between the maximum energy
and the surface magnetic field.  Coupled with observed maximum energies
in pulsars detected by the Compton Gamma-Ray Observatory (CGRO),
this behavior seems to be augmented by an
apparent decline of emission altitude with \teq{B_0}.  Such a trend,
which is not anticipated in outer gap models, is
a distinctive characteristic that can be probed by NASA's upcoming 
Gamma-Ray Large Area Space Telescope (GLAST, to
be launched in mid-2008).  This core NASA astrophysics mission is expected
to add at least 50-100 pulsars to the extant database, possibly many more,
perhaps around half of which will offer clean \teq{\emax} determinations
to refine Figure~\ref{fig:hecutoff} further.  The maximum
energy is generally in the 1--10 GeV band for normal young pulsars, but
can be much lower \cite{hbg97,bh01} for highly magnetized ones, and
also much higher for millisecond pulsars, so that signals in the
30--100 GeV band are possible \cite{brd00,hum05} 
for polar cap models via synchrotron/curvature cascades if the
field is low enough.  

Such a statistical correlation for a population of pulsars is not the only signature
of the operation of \teq{\gamma\to e^+e^-}.  Normally, maximum energy 
turnovers resulting from a moderately energy-dependent attenuation of photons in 
astronomical sources assume an exponential form 
\teq{dn_{\gamma}/d\erg_{\gamma}\propto \exp \{ -\erg_{\gamma}/\erg_{\rm cutoff}\} }.
In this particular instance, the attenuation process is exponentially-sensitive
[as evinced in Eq.~(\ref{eq:pp_asymp_rate})] to the incoming photon energy
for subcritical field strengths; a similarly rapid onset should be experienced
when \teq{B\gtrsim 1}.  The upshot is that magnetic pair creation should impose
a {\it super-exponential} cutoff in pulsar gamma-ray spectra of the approximate form
\teq{dn_{\gamma}/d\erg_{\gamma}\propto \exp \{ -\alpha \exp [-\emax /\erg_{\gamma} ]\} }
for some constant \teq{\alpha}.  The severity of the cutoff will depend on the range of
altitudes and magnetic latitudes sampled in the emission region, but the net
result is typically \cite{dh82,dh96} a sharp turnover that is distinguishable from
those generated by other attenuation processes.  It is anticipated that GLAST
will clearly be able to discern such super-exponential cutoffs, if they are present,
in a number of pulsars, and accordingly offer discrimination between the
viability of polar cap and outer gap models for high energy pulsar emission \cite{rh07}.

The presence of strong fields virtually guarantees a strong
polarization signal in polar cap models, and when these couple with
spectral structure and temporal information, particularly powerful
observational diagnostics are possible.  Due to the polarization-dependence
of the pair production rates, this may provide a potentially fruitful tool,
particularly for highly-magnetized pulsars with \teq{B\gtrsim 10^{13}}Gauss, since the
attenuation cutoffs fall in the 1-10 MeV band.  Hard gamma-ray 
pair production tracking experiments
like GLAST are generally not afforded the opportunity to act as
polarimeters, being limited by multiple scattering in tracking chambers above
300 MeV.  Medium energy gamma-ray experiments, on the other hand, are
ideally suited to polarization studies, essentially via their sampling of Compton
scattering kinematics.  Gamma-ray polarimetry is no longer a distant
dream \cite{ldh97}.  Europe's current
INTEGRAL mission, and NASA's solar RHESSI telescope have some ability to 
detect polarization in bright sources
such as gamma-ray bursts \cite{cb03,kalem07} and X-class solar flares.   
Moreover, polarimetric capability 
in the hard X-ray and soft gamma-ray bands is a high priority for next-generation 
tracking Compton detectors such as the Advanced Compton Telescope 
experimental initiative \cite{boggs06}.

As an interesting aside pertaining to a completely different astrophysical
context, it is worth noting that opacities for pair creation are currently
included in simulation codes for the propagation of ultra-high energy
cosmic rays and air shower initiation in the Earth's upper atmosphere
and magnetosphere.  This presumes a possibility that these energetic
particles could be photons, something that is currently not a popular
perception, but has not yet conclusively been excluded.  For completeness,
the shower simulations for ground-based detector arrays (e.g. \cite{risse04,risse05})
such as Fly's Eye/HIRES, AGASA and Auger include \teq{\gamma\to e^+e^-} 
rates when treating photon primaries, which for
terrestrial fields of \teq{B\sim 1-10}Gauss sample the asymptotic 
regime in Eq.~(\ref{eq:pp_asymp_rate}).  Any photons in the \teq{\erg_{\gamma}
\sim 10^{20}}eV domain can effectively convert into pairs off the geomagnetic field
above the ionosphere and thereby instigate a so-called ``pre-shower'' that is
subsequently reprocessed in a complicated hadronic cascade in the
atmosphere below.  Such a convolution of interactions precludes definitive determination
of the action of \teq{\gamma\to e^+e^-} even if a photon primary origin were
eventually concluded.  Yet a remarkable aspect of this field is that the 
participating physicists attribute much greater certainty to the strong-field QED physics than
to extrapolations of \teq{B=0} photonuclear cross sections to such extreme energies.

\subsection{Photon Splitting}
 \label{sec:split_obs}

It is clear from the rates displayed in Figure~\ref{fig:pprod_split} that the necessary
requirement for photon splitting to operate in pulsar magnetospheres is that
\teq{B\gtrsim 1}.  This turns the focus to
a small subset of known neutron stars/pulsars that possess ultrastrong fields,
termed {\it magnetars}.  Observations at X-ray energies have
yielded detections of both long periods and high period derivatives in
two types of such sources, anomalous X-ray pulsars (AXPs) and soft
$\gamma$-ray repeaters (SGRs), which suggest dipole spin-down fields in
the range \teq{10^{14} - 10^{15}} Gauss.  The AXPs are a group of seven
or eight pulsating X-ray sources with periods in the range 6--12
seconds, and are continuously spinning down \cite{vg97}.  
The SGRs are a type of $\gamma$-ray transient source that
undergoes repeated bursts, some of them enormous (e.g. \cite{hurl99}); 
four are definitively known to exist. 
With the possible exception of the unconfirmed reports \cite{shitov00}
of a low frequency pulsed
detection of a counterpart (PSR J1907+0919 at 111 MHz) to the soft
gamma repeater SGR 1900+14, 
none of these sources has detectable pulsed radio emission.  It is notable,
however, that an unusual related source, the transient AXP XTE J1810-197,
has exhibited transient and peculiar pulsed radio emission \cite{camilo06}.

\begin{figure}[ht]
   \begin{center}
   \includegraphics[width=5.27truein]{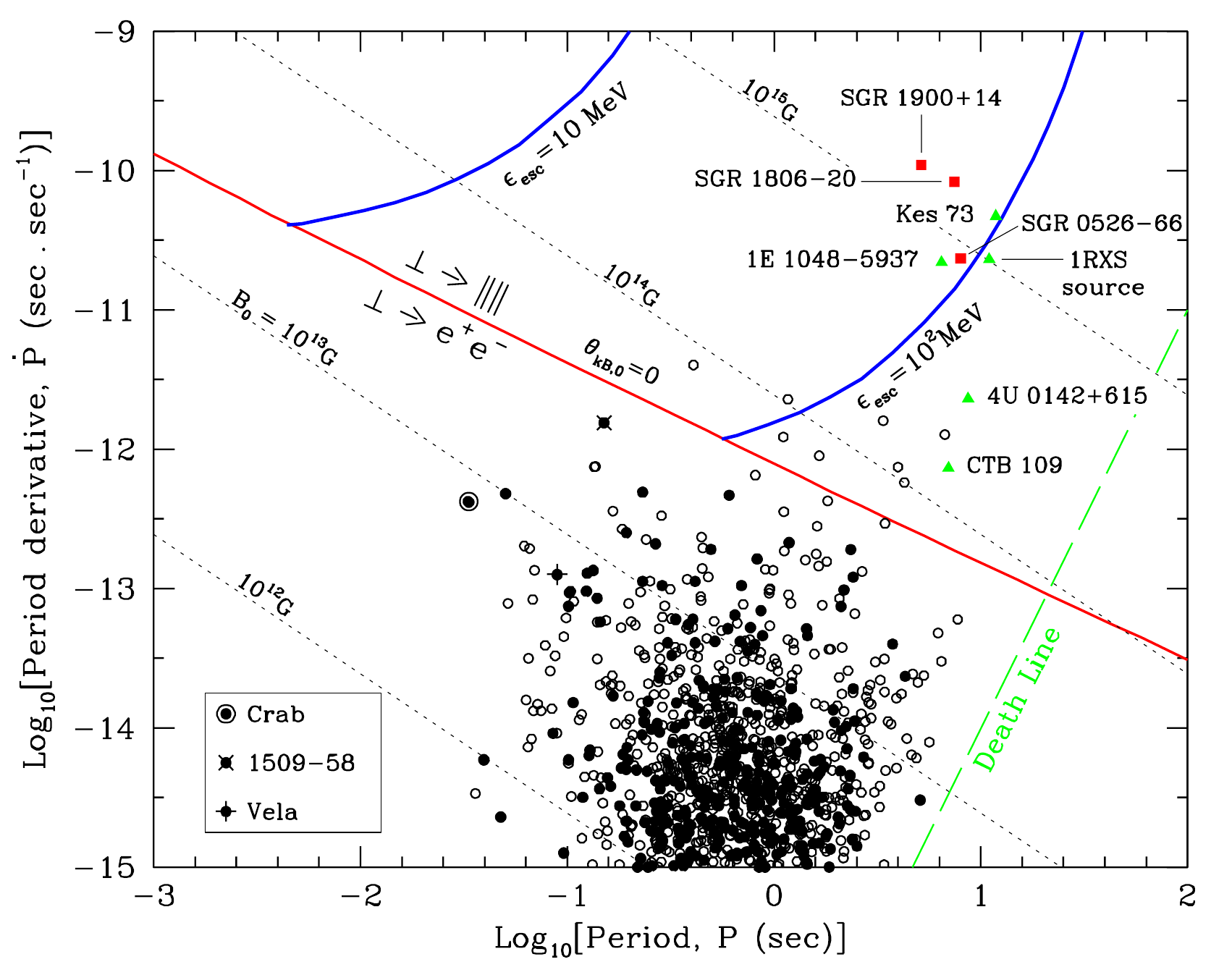}
   \end{center}
   \vspace{-25pt}
\caption{The upper portion of the \teq{P}-\teq{\dot{P}} diagram, with filled circles denoting the
locations of 541 members of the Princeton Pulsar Catalogue \cite{tml93}, and
open circles marking 1599 radio pulsars in the more recent ATNF Pulsar Catalogue
\cite{atnf05} (only those with \teq{\dot{P} \geq 10^{-15}} are visible).  
The approximate boundary (diagonal red line) demarcating 
when splitting dominates (uppermost \teq{{\dot P}}) and when pair creation 
is prolific (lower \teq{{\dot P}} for conventional pulsars) is indicated, being 
computed in \cite{bh01}. Above this,
the heavy-weight curves labelled \teq{\erg_{\rm esc}=10}MeV and
\teq{\erg_{\rm esc}=10^2}MeV are contours for the escape energy of
photon splitting, \teq{\perp\to\parallel\parallel};
to the right of these, the magnetosphere is
transparent to 10 MeV and 100 MeV photons, respectively.
}
 \label{fig:PPdot_split} 
\end{figure}

A central principal of the radio pulsar paradigm is that the emission
is coupled to a prolific presence of pairs \cite{sturr71,rs75}.  As pulsars age,
their periods lengthen, due to the action of magnetic dipole radiation torques.
Concomitantly, the geometrical confines for their acceleration
and emission locales on open field lines become more constricted,
roughly as the polar cap opening angle \teq{\Theta_{\rm cap} \approx 
[2\pi R_{\rm NS}/(Pc)]^{1/2}} shrinks.  This establishes a beaming of radiation
generated by accelerated pairs, and the collimation angles \teq{\theta_{\rm kB}}
for potentially pair producing photons drop.  Eventually, the traversal distance in the magnetosphere required to cross pair threshold exceeds the
scale for magnetic field decline (i.e. \teq{\sim R_{\rm NS}}).  Pair
creation then shuts off, and presumably so does the radio emission, i.e. radio
pulsars ``die.''  This has led to the concept of a pulsar {\it death line} at
long periods, something that is illustrated in Figure~\ref{fig:PPdot_split},
which is the \teq{P}-\teq{\dot{P}} diagram, the conventional depiction of pulsar 
phase space.  The location of this boundary depends on observables \teq{P}
and the period time derivative \teq{{\dot P}}, 
through the relation for the polar field \teq{B_0} that is inferred
for magnetic dipole torques seeding rotational deceleration (e.g. \cite{bh01}):
\begin{equation}
   B_0 \; =\; 6.4\times 10^{19} [P\, {\dot P}]^{1/2} \; \hbox{Gauss}\quad .
 \label{eq:spindownB}
\end{equation}
Dotted diagonal lines in
Figure~\ref{fig:PPdot_split} denote contours of constant field strength.
The locations of over 1500 observed radio pulsars are marked in this Figure.
Three of the 8 known gamma-ray pulsars, namely the 
Crab, Vela and PSR 1509-58 are also highlighted (see lower inset), 
together with the positions of five radio-quiet anomalous X-ray pulsars
(filled green triangles) and SGRs 0526-66, 1806-20 and 1900+14 (filled red squares) 
in the upper right, constituting the sub-class of magnetars.

An alternative mechanism for suppression of radio emission, specifically in high field pulsars 
and magnetars, was proposed by \cite{bh98,bh01}.  
For local fields \teq{B\gtrsim 1}, photon splitting can be an effective competitor
to pair creation.  This arises principally because the pair threshold is
crossed only after emitted photons are transported through the magnetosphere to acquire
significant angles with respect to field lines.  During this time,
when \teq{B\gg 1}, photon splitting can have time to act, and its probability
of attenuating the photons and pushing them further below pair threshold
by energy degradation is sensitive to the local field strength \teq{B} and the 
polar cap geometry, i.e \teq{\Theta_{\rm cap}}.  Consequently, the competition
between splittings \teq{\perp\to\parallel\parallel} and pair conversions
\teq{\perp\to e^+e^-} is defined by the principal pulsar observables
\teq{P} and \teq{{\dot P}}.  Detailed discussions and computations of this
competition are presented in \cite{bh98,bh01}.  They include
all the features of QED discussed above, and the
general relativistic distortions of the magnetosphere and the photon trajectories 
in a Schwarzschild geometry.  The principal result is an identification of the
approximate \teq{P} and \teq{{\dot P}} pulsar phase space when it is
anticipated that photon splitting will dominate pair creation for \teq{\perp}
photons.   This is depicted in Figure~\ref{fig:PPdot_split}.  Since pair
suppression is putatively identified with a lack of radio emission in pulsars,
this defines regimes where photon splitting might seed {\it radio quiescence}
in SGRs and AXPs.  Clearly there are some recently-discovered pulsars above the
boundary, a situation contrasting the complete demarcation highlighted in \cite{bh98}
that pertained to the Princeton Catalogue era.

The placement of the quiescence boundary in the Figure presumes that
the emission region is very near the stellar surface, a criterion that was specifically
adopted when identifying the boundaries in Figure~\ref{fig:PPdot_split}.
If the altitude \teq{R_0} of the photon attenuation locale moves above the
surface, then the putative radio quiescence boundary correspondingly moves to
higher \teq{{\dot P}}, i.e. higher surface polar fields \teq{B_0}.  Since the local
dipole field scales as \teq{B\sim B_0 (R_0/R_{\rm NS})^{-3}}, which is pinned
by the balanced competition between pair creation and splitting, and the inferred
value of \teq{B_0} scales with observables as \teq{B_0\propto ({\dot P}\, P)^{1/2}},
the location of the quiescence boundary in the \teq{P}-\teq{\dot{P}} diagram
approximately obeys \teq{{\dot P}\propto R_0^{+6}} for fixed pulsar periods \teq{P}
(and therefore also fixed polar cap angular sizes).  
Hence, small increases in \teq{R_0} above \teq{R_{\rm NS}} can easily
\cite{bh01} move the boundary above the radio pulsar population
displayed in the Figure.  It should be emphasized that if Adler's
splitting selection rules \cite{adler71} apply here, then photons of \teq{\parallel} polarization
are only subject to pair creation.   This implies only a partial suppression
of \teq{\gamma\to e^+e^-} by splitting, which may lead to a postponement of
pair creation eventually to higher altitudes \cite{bh01}.  However, 
photons of \teq{\perp} polarization dominate the photon population
generated by the principal primary and secondary emission processes of
polar cap models for $\gamma$-ray pulsars \cite{dh82,dh96},
namely cyclotron/synchrotron radiation, curvature emission and resonant
Compton scattering.  Hence, the partial suppression of pair creation by photon splittings
\teq{\perp\to\parallel\parallel} is in fact likely to be quite effective at typical 
magnetar field strengths.

While this identifies a signature and observable action of photon splitting in
an astrophysical setting, the connection to radio emission is the weak link
in this logical construct.   The commonly-held belief that the presence of
a profusion of pairs is requisite for strong radio emission is not yet proven.
The actual nature of the coherent radio emission process(es) is still unknown,
a major problem in the study of pulsars.  If the density of pairs is not relevant
for the production of bright radio emission, then the action of splitting in
magnetar field regimes has no formal connection to the observed radio
quiescence or faintness of SGRs and AXPs.  Hence this property cannot,
at present, be regarded as a definitive measure of the action of splitting in
pulsars.  For potentially more direct evidence for the operation of
\teq{\gamma\to\gamma\gamma} in pulsars, one turns again to spectroscopic
evidence similar to that discussed in Section~\ref{sec:pprod_obs} for
pair creation.  Noting the maximum energy
contours in  Figure~\ref{fig:PPdot_split}, for magnetars and pulsars with fields
above \teq{4\times 10^{13}}Gauss, photon splitting
should prohibit any emission above \teq{\sim 100}MeV, though prominent
signals below 100 MeV are possible \cite{bh07} due to
the efficiency of resonant Compton upscattering of surface thermal X-ray 
photons.  Accordingly, photon splitting should generate the spectral turnovers
in supercritical fields, not solely pair creation, and the physics of the process
guarantees that the cutoff energies are polarization-dependent.  This effect
was discussed in \cite{hbg97}, where a focal study of the high-field pulsar
PSR 1509-58 indicated that the observed spectral data in the sub-100 MeV
range implied a cutoff that could only be modelled in a polar cap scenario
if photon splitting was acting in addition to general relativistic effects.  The
data for this pulsar do not permit much play in the parameter space, so the
inference that splitting is active in PSR 1509-58 is quite strong.  Yet it is
crucial to find one or more other high-\teq{B} pulsars where such constrained
inferences can be made.  Unfortunately such a prospect is not yet on the horizon.
While NASA's imminent GLAST mission will discover an array of new gamma-ray
pulsars, this issue can only be addressed by a sensitive, next-generation 
telescope in the 1--30 MeV range such as the Advanced Compton Telescope.
Yet this specific science question, acquiring strong evidence for the action
of an untested prediction of strong-field QED, forms part of the many motivations
for developing a new medium-energy gamma-ray mission that has
polarimetric capability.

\section{Conclusion}

Two of the exotic and fascinating predictions of quantum electrodynamics
in strong magnetic fields have been highlighted in this paper, one-photon
pair creation and photon splitting.  It is an impressive indicator of the
confidence placed in the theory of QED by physicists that, in spite of
a lack of experimental vindication in terrestrial laboratories, they are both
calculated theoretically in spite of significant degrees of mathematical 
complexity (particularly for photon splitting), 
and anticipated almost routinely in the astrophysical context of pulsars and 
elsewhere.  There have been various limited attempts to explore these processes
in laboratories, sometimes indirectly by searching for birefringence signatures
using lasers in optical cavities \cite{cam93,zava07}, but so far without 
definitive results.  Largely this is because the
required fields are beyond present experimental capabilities: for
\teq{B\sim 1}MGauss, super-TeV energy photons are required to start to
approach the criterion \teq{\omega B\gtrsim 0.1} (in dimensionless units) for
\teq{\gamma\to e^+e^-} to become significant.  It is not beyond reason that
building from the status quo of current powerful terrestrial magnets and
particle accelerator technology such as the Large Hadron Collider (LHC),
it may prove possible to access the appropriate \teq{\omega}-\teq{B} phase 
space in a laboratory setting in the coming decades.  In the meantime, it
is of great interest to explore their application to astrophysics, not just to anticipate
an alternative proof of the action of these QED processes, but also to
enhance our understanding of the exotic cosmic sources, specifically pulsars,
that can provide a conducive environment for the activity of \teq{\gamma\to e^+e^-}
and \teq{\gamma\to\gamma\gamma}.

\noindent\textbf{\\Acknowledgements  }
I thank Dr. Alice Harding and Prof. Don Melrose for many collaborative insights
over the years relating to QED processes in strong magnetic fields, and
Dr. Ken Hines (deceased) for starting me on my way in this field.

\end{document}